\documentstyle{article}
\begin{document}
\baselineskip=0.3in

\hspace{9cm}  ITP-SB-92-74

\hspace{9cm} December 22, 1992
\vspace{0.6in}
\begin{center}
{\large\bf
Numerical Study of Periodic Instanton Configurations in Two-dimensional
       Abelian Higgs Theory}

\vspace{0.3in}
V.V.~Matveev\footnote{E-mail address: VMATVEEV@max.physics.sunysb.edu}

{\it
   Department of Physics, State University of New York at Stony Brook,

   Stony Brook, NY 11794-3800, USA

{and}

   Institute for Nuclear Research of the Academy of Sciences of Russia,

   60th October Anniversary Prospect, 7a, Moscow 117312, Russia
}
\end{center}

\vspace{0.3in}
\begin{abstract}
  Numerical minimization of the Euclidean action of the two-dimensional
Abelian Higgs model is used to construct periodic instantons, the euclidean
field configurations with two turning points describing transitions between
the vicinities of topologically distinct vacua. Periodic instantons are found
at any energy ( up to the sphaleron energy $E_{sph}$ ) and for wide range of
parameters of the theory. We obtain the dependence of the action and the
energy of periodic instanton on its period; these quantities directly
determine the probability of certain multiparticle scattering events.

\end{abstract}
\newpage

{\large\bf 1.} In the four-dimensional standard electroweak
theory, instanton-induced transitions
account for interesting phenomenon of baryon- and lepton-number violation;
here we consider a much simpler two-dimensional Abelian Higgs theory
which is analogous to the former to the extent that it also possesses a
non-trivial vacuum structure. In both theories, one may anticipate the
existence of periodic instanton configurations, which are solutions to the
euclidean field equations with two turning points and zero winding number
[1] (see also [2,3]). As it is shown by S.Khlebnikov et al [1],
 periodic instanton is an exact saddle point in the functional integral for
the probability of the multiparticle scattering event that leads to a
transition between the vicinities of topologically distinct vacua. It is also
shown that the analytic continuation of the periodic instanton to the
Minkowski domain through its turning points provides the most probable
initial and final states of this transition. In other words, the transition
induced by the periodic instanton has the largest probability at a given
energy. In the semiclassical approximation, this probability was found to
be [1]
\begin{equation}
  \sigma_E=\exp\{ - S + ET \}
\end{equation}
where T is the period and S is the action per period of the periodic
instanton of energy~E.

At low energies, the periodic instanton configuration can be
approximated by a temporal alternating sequence of instantons and
anti-instantons separated by equal intervals of $T/2$.
The period T grows exponentially as the energy approaches zero.

At energies close to the sphaleron [4] energy $E_{sph}$, the periodic
instanton can be described as the sum of the static sphaleron configuration
and the oscillation in the negative eigenmode around the sphaleron [5,6].
As the energy $E$ approaches $E_{sph}$, the amplitude of this oscillation
vanishes. The period of the periodic instanton in this approximation is
equal to the period of this eigenmode.

  At arbitrary energies, the analytic form of the periodic instanton is not
known; furthermore, the boundary conditions ( fields at the turning points )
are also unknown. In the present paper we evaluate numerically the periodic
instantons of the two-dimensional Abelian Higgs theory at any energy and
for wide range of parameters of the theory. We confirm the expected
properties of these configurations and show, in particular, that the
semiclassical probability $\sigma_E$, equation (1), interpolates between
its instanton value and unity as the energy varies from zero to $E_{sph}$.
\vspace{0.3in}

{\large\bf 2.} The Lagrangian of the two-dimensional Abelian Higgs model
reads
$$
 L=-\frac{1}{4}F^2_{\mu\nu}+\left|(\partial_\mu-igA_\mu)\phi\right|^2-\lambda
\left(|\phi|^2-\frac{v^2}{2}\right)^2 \qquad \mu, \nu = 0, 1.
$$
Here $\phi$ is the complex Higgs field and $A_\mu$ is the U(1) gauge field.
The particle spectrum of this model consists of the Higgs boson of mass
$M^2_H=2\lambda v^2$ and the vector boson of mass $M^2_W=g^2 v^2$. Let us
consider the temporal gauge $ A_0=0 $; the remaining symmetry we will fix
according to the condition
$$
A_1(x,t=t_0)\ =\ 0.
$$
Later we will define the value of $ t_0 $.

In this gauge we may choose the following set of topologically distinct
vacua
$$
\phi _n(x)=v\ e^{i\alpha _n(x)},\qquad A_1^n(x)=\frac{1}{g}\partial _1\alpha
_n,
$$
$$
(\alpha_n(x=+\infty)-\alpha_n(x=-\infty))=2\pi(n+\frac{1}{2}),
$$
$$
 \phi_n(-\infty)=-v,\quad \phi_n(+\infty)=v.
$$
This set is related to a conventional one with $ \alpha _n(x=+\infty)-
\alpha _n(x=-\infty)=2\pi n\ $ by a time-independent gauge transformation
that plays no role in any physical process.

In the space of field configurations, the neighboring vacua are separated
by a static energy barrier. The top of this barrier is associated with the
sphaleron [4], which is the static unstable solution to the field
equations.
In our gauge this solution is the ordinary kink (cf. [5,7])
$$
A^{sph}_\mu=0,\quad \phi^{sph}=\frac{v}{\sqrt{2}} \tanh \frac{M_H x}{2}.
$$
Its energy ( the height of the barrier ) is equal to
$$
E_{sph}=\frac{\sqrt{8\lambda}\ v^3}{3}.
$$

Among the modes of oscillations of fields around the sphaleron there is one
negative eigenmode, which makes the sphaleron unstable. The frequency of
this mode is [6],
\begin{equation}
\omega^2_- = - \frac{M_H^2}{8} \left( \sqrt{1+\frac{16M_W^2}{M_H^2}}+1
\right).
\end{equation}
This quantity determines the minimal period of a periodic instanton
configuration: as energy increases from zero to $E_{sph}$ the period changes
from infinity to $T_-=2\pi/|\omega_-|$.

  The instanton solution of this model is the well-known
Abrikosov-Nielsen-Olesen
vortex [8,9]. The instanton is a zero-energy, finite action
euclidean solution to the field equations interpolating between two
neighboring vacua. Its winding number is equal to~1,
$$
 \left.\frac{1}{2\pi}\oint A_\mu^I\ dx_\mu\ =\ \frac{1}{2\pi}
\int A_1^I\ dx^1\right|_{t=+\infty}^{t=-\infty}\ =\ 1.
$$

At various values of $M_H/M_W$ this solution was obtained
numerically in ref. [10].

\vspace{0.3in}
{\large\bf 3.} We construct the periodic instanton field for various values
of the
period (ranging from $T_-$ to infinity) using the numerical minimization
of the euclidean action. Thus, the values of energy and action of the
periodic instanton and its other properties are obtained as functions of its
period.

The minimization is performed on the space of fields satisfying the following
conditions
\begin{equation}
\dot A_1(x,-\frac{T}{2}) = \dot A_1(x,0) = 0, \quad A_1(-\infty,t) =
A_1(+\infty,t) = 0,
\end{equation}
\begin{equation}
\dot \phi(x,-\frac{T}{2}) = \dot\phi(x,0) = 0,\quad \phi(-\infty,t) = -v,
\quad \phi(+\infty,t) = v.
\end{equation}
\begin{equation}
A_1(x,t_0=-\frac{T}{4} ) = 0.
\end{equation}
Notice that Gauss' law $\partial_1\partial_0 A_1 = -\frac{i}{2}g
(\dot\phi \phi^* -\phi \dot\phi^* )\ $ is satisfied at the
 turning points; this provides
that it will be automatically satisfied at all times for the fields
minimizing the action and obeying the field equations.

  Equation (5) specifies our choice of gauge. This particular choice leads
to the following additional symmetries of the solution
$$
A_1(x,-\frac{T}{4}-t) = -A_1(x,-\frac{T}{4}+t),
$$
$$
\phi(x,-\frac{T}{4}-t) = \phi^*(x,-\frac{T}{4}+t).
$$
This symmetry allows one to perform minimization on the time interval equal
to the quarter of the period, which is convenient for the numerical study.

The calculations are performed on a two-dimensional lattice with the time
dimension of $[-\frac{T}{4},-\frac{T}{2}]$; the choice of the spatial
x-dimension depends on the desired accuracy and the values of parameters
$M_H$ and $M_W$.

First, an arbitrary configuration satisfying the conditions (3-5) is chosen.
Then the following step is repeated necessary number of times:

{ }{ } The trial function to be added to one of the fields is chosen as the
product of a random time-dependent trigonometric harmonic and x-dependent
gaussian function of random width and position. Then a standard
single-parameter minimization technique is applied with respect to the
amplitude of this trial function. Along with the trial function its
derivatives of known analytical form are also calculated and added after
minimization to the corresponding derivatives of the field. In this sense
the derivatives are calculated 'exactly': we do not use any difference
approximations. This greatly increases the accuracy of the results.

\pagebreak
{\large\bf 4.} Calculations were performed for different values of the ratio
$M_H/M_W$ ranging from $\frac{1}{4}$ to 4. Here we present the
results mainly for the case $M_H=M_W=1$; the results at other
values of $M_H/M_W$ are similar.

Figs. 1 and 2 show respectively the dependence of the energy and the action
per period of the periodic instanton on its period. In full agreement with
the expectations, the energy approaches $E_{sph}$ as the
period decreases. If the period is set to a value smaller then $T_-$,
calculations lead to a static sphaleron configuration, since there are no
other periodic solutions at $T<T_-$. The numerically obtained minimal value
of the period, $T_-$, coincides with the expected value, determined by eq.
 (2).
At large values of the period the energy
is close to zero, and the action is close to twice the value of the
action of the instanton (~which is equal to $\pi$ when $M_H=M_W=1$~).
Functions S(T) and E(T) satisfy the following relation [1]
\begin{equation}
E(T)=\frac{\partial S(T)}{\partial T}
\end{equation}
which provides an additional check for numerical study.

Fig. 3 shows the behavior of $( S-ET )=-\ln\sigma_E$ as the function of
energy, where $\sigma_E$ is the probability of the instanton-induced
multiparticle scattering event, see eq. (1). At $E=E_{sph}$ the value of
$(S-ET)$ reaches zero, as it should have been expected; this corresponds to
an unsuppressed probability of the transition.

For completeness, in figs. 4-6 we show typical periodic instanton
field configurations.
Here the fields are shown for the case $M_W=1, M_H=\frac{1}{4}$.
We have checked that at large T these configurations do indeed coincide
with the instanton--anti-instanton pair.

\vspace{0.2in}
To conclude, we have verified numerically the existence of the periodic
instantons in two-dimensional Abelian Higgs model. They indeed interpolate
between a widely separated chain of instantons and anti-instantons and the
sphaleron as energy increases from zero to $E_{sph}$ and describe most
favorable tunneling events at each energy. The probability of such event
increases with energy starting from $\exp \{-2S_{inst}\}$ at $E=0$ and
becomes no longer exponentially suppressed at $E=E_{sph}$.
\vspace{0.2in}

Author is grateful to Prof. V.A.~Rubakov for his guidance and
help at all stages of this work and to Dr. P.~Tinyakov for useful
discussions.

\vspace{0.3in}
\begin{center}
\large
\bf
References
\end{center}

\hspace{-1cm}[1] S.Yu.~Khlebnikov, V.A.~Rubakov and P.G.~Tinyakov, Nucl.
 Phys. {\bf B367} (1991)
334.

\hspace{-1cm}[2] K.~Funakubo, S.~Otsuki, K.~Takenaga and F.~Toyoda, Prog.
 Theor. Phys.
  {\bf 87} (1992) 663.

\hspace{-1cm}[3] K.~Funakubo, S.~Otsuki, K.~Takenaga and F.~Toyoda,
 Scattering with
    baryon number violation -- SU(2) gauge-Higgs system, preprint
  SAGA-HE-43, Saga University, July 1992.

\hspace{-1cm}[4] F.R.~Klinkhamer and M.S.~Manton, Phys. Rev. {\bf D30} (1984)
 2212.

\hspace{-1cm}[5] A.I.~Bochkarev and M.E.~Shaposhnikov, Mod. Phys. Lett. {\bf
 A2} (1987) 991.

\hspace{-1cm}[6] A.I.~Bochkarev and G.G.~Tsitsishvili, Phys. Rev. {\bf D40}
 (1989) 1378.

\hspace{-1cm}[7] D.Yu.~Grigoriev and V.A.~Rubakov, Nucl. Phys. {\bf B299}
 (1988) 67.

\hspace{-1cm}[8] A.A.~Abrikosov, Sov. Phys. JETP {\bf 5} (1957) 1174.

\hspace{-1cm}[9] H.B.~Nielsen and P.~Olesen, Nucl. Phys. {\bf B61} (1973) 45.

\hspace{-1cm}[10] E.B.~Bogomol`nyi, Soviet Journal of Nuclear Physics {\bf 24}
(1976) 49.

\newpage
\begin{center}
\large
\bf
Figure Captions
\end{center}
\vspace{1in}

{\hspace{-1cm}\bf Figure 1. } Dependence of the energy of the periodic
 instanton
 on its period.

{\hspace{-1cm}\bf Figure 2. } Dependence of the action per period of the
 periodic instanton on its period.

{\hspace{-1cm}\bf Figure 3. } Dependence of the value $(S-ET)$ on the energy.

{\hspace{-1cm}\bf Figure 4. } Real part of the Higgs field for
 $M_W=1,\ M_H=1/4,\  T=32$.

{\hspace{-1cm}\bf Figure 5. } Imaginary part of the Higgs field for
 $M_W=1,\ M_H=1/4,\ T=32$.

{\hspace{-1cm}\bf Figure 6. } Spatial component of the gauge field
 $A_1(x,t)$ for $M_W=1,\ M_H=1/4,\ T=32$.
\end{document}